\documentclass[a4paper]{article}
\usepackage{graphicx,amsmath,amssymb,color, amsthm, natbib}
\newtheorem{theorem}{Theorem}
\newtheorem{assumption}{Assumption}

\newtheorem{remark}{Remark}
\usepackage{comment}

\usepackage{epsfig,subfigure}
\usepackage{epsfig,float}
\usepackage{natbib}
\usepackage{color}
\usepackage{lipsum} 
\usepackage{multirow}
\usepackage{booktabs}   
\usepackage{pifont}     
\usepackage{array}      
\usepackage{graphicx}   
\usepackage{enumerate}
\usepackage{enumitem}
\usepackage[colorlinks=true,citecolor=blue, linkcolor=blue, urlcolor=blue]{hyperref}
\numberwithin{equation}{section}
\renewcommand{\arraystretch}{1.3} 

\date{}
\allowdisplaybreaks

\title{Variable selection in spatial lag models using the focussed information criterion}

\date{}

\begin{document}


	\author{
		\hspace{0.01\textwidth}
			\parbox[t]{0.25\textwidth}{{Sagar Pandhare}
           \thanks{Indian institute of technology (IIT), Bombay, \ Plaksha Univeristy, Punjab \ Email: pandhare.05@gmail.com}}
		\hspace{0.01\textwidth}
		\parbox[t]{0.25\textwidth}{{Divya Kappara$^+$}
        \thanks{$+$ Corresponding author, Mathematics department, IIT, Bombay, Email: kapparadivya@gmail.com; 
        }
		\hspace{0.01\textwidth}}
		\parbox[t]{0.25\textwidth}{{Siuli Mukhopadhyay}
				           \thanks{Mathematics Department, IIT, Bombay,\ Email: siuli@math.iitb.ac.in
                           \\  Authors 1 and 2 contributed to the conceptualization, design, and execution of the study. Author 3 provided guidance in writing and revising the manuscript for important intellectual content. The first and third author's research is supported by Gates foundation, grant number RD/0122-GATES0A-001-EXP.}
                           }
		}

\maketitle

\begin{abstract}
    \noindent Spatial regression models have a variety of applications in several fields ranging from economics to public health. Typically, it is of interest to select important exogenous predictors of the spatially autocorrelated response variable. In this paper, we propose variable selection in linear spatial lag models by means of the focussed information criterion (FIC). The FIC-based variable selection involves the minimization of the asymptotic risk in the estimation of a certain parametric focus function of interest under potential model misspecification. We systematically investigate the key asymptotics of the maximum likelihood estimators under the sequence of locally perturbed mutually contiguous probability models. Using these results, we obtain the expressions for the bias and the variance of the estimated focus leading to the desired FIC formula. We provide practically useful focus functions that account for various spatial characteristics such as mean response, variability in the estimation and spatial spillover effects. Furthermore, we develop an averaged version of the FIC that incorporates varying covariate levels while evaluating the models. The empirical performance of the proposed methodology is demonstrated through simulations and real data analysis. 
\end{abstract}
\section{Introduction}
Spatial models have a wide range of applications in various fields such as economics, geographical sciences, and epidemiology. Such models are flexible enough to quantify the spatial stochastic dependence in the data while simultaneously addressing the dependence of the spatially varying response on exogenous predictors. 
The spatial lag model (SLM) introduced by \cite{cliff1973spatial} is one of the most widely used regression models for spatial data. The model selection problem pertinent to an SLM typically involves the selection of important predictors of the spatially varying response variable and the selection of a suitable spatial adjacency weight matrix that may adequately capture the spatial stochastic structure of the response variable. A few developments have been made to address the problem of variable selection in spatial regression models. \cite{zhang2018spatial} proposed a model selection approach to identify the optimal spatial adjacency matrix from multiple candidates using a Mallows-type criterion. Here, the model selection primarily focuses on determining the most appropriate spatial weights matrix. \cite{yu2024robust} proposed a robust variable selection with exponential squared loss under a spatially varying coefficient setup. Quite recently, \cite{zhang2024variable} addressed variable selection in nonparametric SLM using a deep learning approach. A more general set-up of spatial stochastic process models with spatial covariance structures has been explored by \cite{huang2007optimal} and \cite{wang2009variable} among others. 
\par In spatial analyses, covariate effects are often heterogeneous, with different factors driving outcomes across regions or hotspots. The existing variable selection methods for spatial models are aimed at selecting a single ``best" model, but may overlook locally important variables. They do not take into account the functional form of the estimands of interest and the bias-variance trade-off in the estimation under potential model misspecification. The selection of covariates in an SLM should ideally reflect spatially relevant characteristics such as response levels at hotspots, spatial variability and spatial spillover. We address this research gap by means of the novel focused information criterion (FIC) proposed by \cite{claeskens2003focused} explicitly takes into account the purpose of model selection through a user-specified parametric focus function. In our case, the model selection pertains to identifying the relevant columns of the design matrix $X$ under a fixed spatial weights matrix W, where the goal is to determine which covariates most strongly influence the response given the spatial neighborhood structure. Hence in an SLM context the FIC offers an alternative by aligning variable selection with a specific parameter of interest  such as the mean outcome within a hotspot, or the effect of a covariate in a given subregion etc. For example, if the focus parameter is the mean outcome within a hotspot, FIC selects the model that provides the most accurate estimate of this mean, taking the spatial structure into account. Similarly, for other parametric focus functions—such as a spatial spillover effect—the covariates included in the model chosen by FIC are those most relevant for accurately capturing the effect.

The FIC-based model selection involves minimization of the risk of the focus function estimated under potentially locally misspecified models. The local misspecification setting proposed by \cite{claeskens2003focused} prevents the bias of the estimands to explode as $n\to\infty$ leading to the tractable asymptotic distributions and analysis of the bias-variance trade-off caused due to model misspecification. Ever since its inception, the idea of focussed inference and model selection has received considerable attention and appreciations. \cite{hansen2005challenges} described
the FIC to be ‘an intriguing challenger to existing model selection methods and deserves
attention and scrutiny’. The original setup of the FIC has been extended for various models and data structures such as semiparametric models \cite{claeskens2007asymptotic}, generalised additive partially linear models \cite{zhang2011focused}, high dimensional regression
(\cite{gueuning2018high}, \cite{pandhare2023robust}), time series modeling (\cite{claeskens2007prediction}, \cite{ramanathan2021focused}),  robust regression modelling (\cite{du2018model}, \cite{pandhare2020robust}). An extensive review of advances in the domain of focussed
statistical modelling and inference can be found in \cite{claeskens2008model}, Chapter 6
and \cite{claeskens2016statistical}.
\par The objective of this paper is to facilitate this cutting-edge FIC toolkit for spatial regression models. We work with the setup of a linear Gaussian spatial lag model with exogenous covariates whose selection is of interest. We systematically investigate the asymptotics underlying the FIC mechanism and derive the desired FIC formula. All these results are used to derive the general FIC formula. We also provide some practically useful focus functions relevant to spatial modeling. Moreover, we develop the averaged version of the FIC that is useful for model selection over a range of covariate levels across different regions. 
\par The rest of the paper is organized as follows. The model setting and the problem of focused variable selection, examples of focus functions along with the asymptotics under local model misspecification are provided in Section \ref{sec2}.  Using the local asymptotics in Section \ref{sec2}, the general spatial FIC is obtained in Section \ref{sec3}. The spatial average FIC is formulated in Section \ref{sec4}. The performance of the proposed methodology is investigated empirically through simulations in Section \ref{sec5}. Section \ref{sec6} is devoted to the application to the Boston Housing data where we use the FIC for selection of covariates of the spatially varying housing prices. Conclusions and future directions are summarized in Section \ref{sec7}. The technical assumptions and proofs of all the theoretical results are provided in the appendices.  
\section{Focused variable selection and local asymptotics} \label{sec2}
\subsection{Model and its likelihood}
Consider a spatial lag model (SLM)
\begin{equation}
    Y=\rho WY+X \beta + \epsilon \label{model},
\end{equation}
where $Y=(y_1,\cdots, y_n)^\top$ is $n\times 1$ vector of responses, $W$ denotes an $n\times n$ spatial adjacency matrix, $X$ is an $ n\times p$ matrix of exogenous covariates, $\rho$ is a real-valued spatial autoregression parameter, $\beta$ is a $p\times 1$ vector of regression coefficients and $\epsilon=\left(\epsilon_{1},\cdots, \epsilon_{n}\right)^{\top}$ is an $N_{n}(0,\sigma^{2} I_{n})$ vector of innovations. Note that the Equation \eqref{model} can be expressed as 
\begin{equation*}
    \left(I_{n}-\rho W\right)Y=X \beta+\epsilon. 
\end{equation*}
Hence, if $\rho$ is known, the MLE of $\beta$ is given as 
\begin{equation*}
    \begin{split}
    \hat{\beta}(\rho)&=\left(X^\top X\right)^{-1}X^\top\left(I_{n}-\rho W\right)Y\\
    &=\left(X^\top X\right)^{-1}X^\top Y-\rho \left(X^\top X\right)^{-1}X^\top WY\\
    &=\hat{\beta}_{R}-\rho \hat{\beta}_{L},
    \end{split}
\end{equation*}
where $\hat{\beta}_{R}$ is the OLS/MLE obtained by regressing $Y$ on $X$, while $\hat{\beta}_{L}$ is the OLS/MLE obtained by regressing $WY$ on $X$. The MLE of $\sigma^{2}$ is given as 
\begin{equation*}
    \begin{split}
       \hat{\sigma}^2\left(\hat{\beta}(\rho),\rho\right)
       &= n^{-1}\left[\left(I_{n}-\rho W\right)Y-X\hat{\beta}(\rho)\right]^\top \left[\left(I_{n}-\rho W\right)Y-X\hat{\beta}(\rho)\right]\\
        &=n^{-1}\left(\hat{e}_{R}-\rho \hat{e}_{L}\right)^{\top}\left(\hat{e}_{R}-\rho \hat{e}_{L}\right),
    \end{split}
\end{equation*}
where $\hat{e}_{R}=Y-X\hat{\beta}_{R}$ and $\hat{e}_{L}=WY-X\hat{\beta}_{L}$ denote the residuals obtained based on the regression of $Y$ on $X$ and $WY$ on $X$ respectively. Once the estimator of $\rho$ is obtained, it can be substituted in $\hat{\beta}(\rho)$ and $\hat{\sigma}^2\left(\hat{\beta}(\rho),\rho\right)$ to get the estimates of $\beta$ and $\sigma^2$. The MLE of $\theta=(\rho,\beta,\sigma^2)$ can be obtained by numerically optimizing the concentrated log-likelihood likelihood function given by
\begin{equation}
\log  L(\theta)_{c}=-\frac{n}{2}-\frac{n}{2}\log~2\pi-\frac{n}{2}\log \hat{\sigma}^2\left(\hat{\beta}(\rho),\rho\right)+\log |\left(I_{n}-\rho W\right)|. \label{likelihood}
\end{equation}
Under mild restrictions on the structure of $W$ and $X$, the MLE can be shown to be asymptotically efficient. These assumptions and related discussions are provided in Appendix \ref{App:A}. We refer to \cite{ord1975estimation} and \cite{anselin1988spatial} for detailed discussions on the MLE for spatial models.
\subsection{The focused variable selection problem} 
\par It is of interest to select the important features in the design matrix $X$ in the presence of spatial autocorrelation in the response $Y$. We assume that the largest (wide) model under consideration has all $p+2$ parameters while the smallest (narrow) model is a purely spatial autoregressive model with no exogenous covariates. We assume that all the candidate models lie between the narrow and the wide model. Let $\mathcal{S}\subseteq\{1,2,...,p\}$ denote an index for a subset model with the cardinality $|\mathcal{S}|$. The wide model reduces to a certain subset model when some of the elements in the coefficient vector $\beta$ are set to 0. The case when all elements of $\beta$ are 0 reduces the wide model to the smallest, narrow model, which in our case is the simple spatial lag model. Let $\Pi_{\mathcal{S}}$ denote the $|\mathcal{S}|\times p$ projection matrix such that $\beta_{\mathcal{S}}=\Pi_{\mathcal{S}}
\beta$ denotes the sub-vector of non-zero elements of $\beta$ relevant to a subset model $\mathcal{S}$. 
For instance, suppose $p=4$. Hence $\beta=(\beta_{1},\beta_{2},\beta_{3},\beta_{4})$ is a coefficient vector in the wide model. Suppose a subset model has $(\beta_{1},\beta_{3})$ as coefficients. In this case, $\mathcal{S}=\{1,3\}$ while $\Pi_{\mathcal{S}}$ is given as 

\[
\Pi_{\mathcal{S}}=\begin{pmatrix}
1 & 0 & 0 & 0 \\
0 & 0 & 1 & 0 \\
\end{pmatrix}.
\]
Let $\mu_{W}:\Theta\to\mathbb{R}^{k}$, $1\leq k<n$ denote a vector parametric focus function for a given $W$. We assume throughout that $\mu_{W}$ is continuously differentiable on $\Theta$ for any given $W$. Having specified the spatial weights, the focused variable selection problem in the present context involves the identification of the optimal subset model $\mathcal{S}$ for which the limiting mean squared error of $n^{1/2}(\mu_{W}(\hat{\theta}_{\mathcal{S}})-\mu_{W}(\theta_{\mathcal{S}}))$ is minimum when the models under consideration are potentially misspecified. Here, $\hat{\theta}_{\mathcal{S}}$ is the MLE of $\theta_{\mathcal{S}}=(\rho,\beta_{\mathcal{S}},\sigma^{2})^{\top}$ under subset model $S$. The set up of a vector valued function enables us to determine a subset of variables that is optimal for various focus functions simultaneously. Moreover, we let the underlying focus function depend on $W$ in general so that spatial connectivity is also incorporated in model selection. 
\subsection{Examples of focus functions}
\begin{enumerate}
 \item Conditional expectation given neighbors: 
A common choice of focus function in regression-type models is the 
linear predictor. In the SLM framework, 
this corresponds to the conditional mean response at location $i$, 
given its neighbors:  
\begin{equation}
\mu_{W,i}(\theta)  
= \rho \sum_{j=1}^{n} w_{ij} y_j + x_i^{\top}\beta,
\label{focus1}
\end{equation}

 where $w_{ij}$ denotes the $(i,j)$-th element of $W$ and, $x_{i}^{\top}$ is the $i$-th row of $X$, $i,j=1,2,...,n$. This focus function is based on the mean response at a fixed spatial location such as hotspots. The FIC using this focus function incorporates the localized effect of spatial regions in variable selection.
\item  Variability in the parameter estimates: 
In principle, it is recommended to incorporate the variability in the estimators while selecting the models using the FIC. Under the standard regularity conditions such as those in \cite{ord1975estimation}, the inverse Fisher information $\mathcal{I}(\theta)^{-1}$ is the asymptotic covariance matrix of the MLE of $\theta$. Since the maximum eigen value reflects the highest variability in the parameter estimates, we choose the focus function to be 
\begin{equation}
\mu_{W}(\theta)=\lambda_{W}(\theta), \label{focus3}
\end{equation}
where $\lambda_{W}(\theta)$ is the maximum eigenvalue of $\mathcal{I}(\theta)^{-1}$ for a given choice of $W$. 
  \item     Regression coefficients: It may be of interest to select a suitable subset model by evaluating the impact of exogenous covariates on the response. In such a case, the ideal choice of a focus function is 
   \begin{equation}
   \mu_{W}(\theta)=\beta. \label{focus2}
\end{equation}
   \item   Spatial spillover effect along with the model parameters: The term $\log |I_{n}-\rho W|$ is crucial in the likelihood \eqref{likelihood} of an SLM model since it addresses the spatial spillover effect which refers to the impact of an event occurring in one spatial unit on its neighboring regions. When $|I_{n}-\rho W|=1$ (i.e. $\rho=0$), the likelihood function reduces to that of the classical regression model. This term augmented with the regression coefficients and error variance represents the spatial dependence, the impact of covariates and the variability in the response. With this view, we choose the  focus function as 

\begin{equation}
   \mu_{W}(\theta)=\left(\log |I_{n}-\rho W|,\sigma^{2},\beta \right). \label{focus4}
\end{equation}

\end{enumerate}
\subsection{Local model misspecification and local asymptotics}\label{sec2.3}
While selecting a suitable model using the FIC, it is essential to utilize the bias-variance trade-off in the estimation under a potential model misspecification. Adding unimportant variables may reduce the bias, but will increase the variance of the estimation (\cite{hastie2005elements}), Chapter 7.3. However, eliminating some of the important variables will increase the bias in the estimation at the expense of variability. In order to tract the bias and the variance of the MLE under model misspecifcation, it is assumed that ``least false parameter" $\theta$ lies in an open sphere around the null value $\theta_{0}$ such that the radius of the sphere shrinks at the rate $o(n^{-1/2})$ . The well-known Le Cam's contiguity lemmas \cite{lecam1970assumptions} can then be employed to obtain the asymptotic distribution of the MLE under a sequence of locally perturbed probability models. The detailed discussions on the local misspecification and local asymptotics with applications to focused variable selection can be found in \cite{claeskens2008model}, Chapter 5.   
\par In the present context, since $\rho$ and $\sigma^{2}$ appear in all the models, they are left unperturbed while it is assumed that  there exists a sequence  $\{\beta_{n},n\geq 1\}$ of regression coefficients and a non-zero constant vector $\delta$ in $\mathbb{R}^{p}$ such that 
\begin{equation}
\beta_{n}-\beta_{0}=o\left(n^{-1/2}\delta\right). \label{2.6}
\end{equation}
The parameter $\delta$ controls the bias in the estimation under model misspecification leading to the root-$n$-consistent MLE with the asymptotic bias of order $O(\delta)$. For all practical purposes, we set the null $\beta_{0}$ to 0 (without loss of generality), which corresponds to the case where a model reduces to the smallest model consisting of only $\rho$ and $\sigma^{2}$. 
\par Partition the Fisher information $\mathcal{I}(\theta)$ according to $(\rho,\sigma^{2},\beta)$ and in that order as 

\[
\mathcal{I}(\theta)=\begin{pmatrix}
\mathcal{I}_{\rho,\rho}  & \mathcal{I}_{\rho,\sigma^{2}} & \mathcal{I}_{\rho,\beta} \\
\mathcal{I}_{\sigma^{2},\rho} & \mathcal{I}_{\sigma^{2},\sigma^{2}}  & \mathcal{I}_{\sigma^{2},\beta}  \\
\mathcal{I}_{\beta,\rho}  & \mathcal{I}_{\beta,\sigma^{2}} & \mathcal{I}_{\beta,\beta}
\end{pmatrix}.
\]
 It can be seen that since $\beta$ and $\sigma^{2}$ are orthogonal, $\mathcal{I}_{{\beta},\sigma^{2}}$ and $\mathcal{I}_{\sigma^{2},\beta}$ are both zero matrices of orders $p\times 1$ and $1\times p$ respectively. The $(|\mathcal{S}|+2)\times (|\mathcal{S}|+2)$ submatrix of the above Fisher information matrix corresponding to a subset model $\mathcal{S}$ is given as 
\begin{equation*}
\mathcal{I}_{\mathcal{S}}=\mathcal{I}(\theta_{\mathcal{S}})=\begin{pmatrix}
\mathcal{I}_{\rho,\rho} & \mathcal{I}_{\rho,\sigma^{2}}  & \mathcal{I}_{\rho,\beta}\Pi_{\mathcal{S}}^{\top} \\
\mathcal{I}_{\sigma^{2},\rho}& \mathcal{I}_{\sigma^{2},\sigma^{2}}  & O_{1\times |\mathcal{S}|}\\
\Pi_{\mathcal{S}}\mathcal{I}_{\beta,\rho} & O_{|\mathcal{S}|\times 1} & \Pi_{\mathcal{S}}I_{\beta,\beta} \Pi_{\mathcal{S}}^{\top} 
\end{pmatrix}.
\end{equation*}
The blockwise inverse of the above partitioned matrix is represented as 
\[
\mathcal{I}_{\mathcal{S}}^{-1}=\begin{pmatrix}
\mathcal{I}_{\mathcal{S}}^{\rho,\rho} & \mathcal{I}_{\mathcal{S}}^{\rho,\sigma^{2}} & \mathcal{I}_{\mathcal{S}}^{\rho,\beta}\\
\mathcal{I}_{\mathcal{S}}^{\sigma^{2},\rho} & \mathcal{I}_{\mathcal{S}}^{\sigma^{2},\sigma^{2}} & O_{1\times|\mathcal{S}|} \\
\mathcal{I}_{\mathcal{S}}^{\beta,\rho} & O_{|\mathcal{S}|\times 1}  & \mathcal{I}_{\mathcal{S}}^{\beta,\beta}
\end{pmatrix}.
\]
In what follows, $U_{n,\rho}$, $U_{n,\sigma^2}$ and $U_{n,\beta}$ denote the gradients of the log-likelihood function \eqref{likelihood} with respect to $\rho$, $\sigma^2$ and $\beta$ respectively. Let $U_{n,\mathcal{S},\beta}=\Pi_{\mathcal{S}}U_{n,\beta}$ denote the subvector of $U_{n,\beta}$ corresponding to a subset model $\mathcal{S}$. The following theorem provides the asymptotic distributions under the local misspecification.
\begin{theorem}\label{Th:1}
Under the assumptions in the Appendix \ref{App:A} and local misspecification setting given by Equation (\ref{2.6}), it holds for every $\mathcal{S}\subseteq\{1,2,\cdots,p\}$ that 
\begin{enumerate}
 \item  
 \begin{equation*}
n^{-1/2}
\begin{pmatrix}
U_{n,\rho}\\
U_{n,\sigma^2}\\ U_{n,\mathcal{S},\beta}
\end{pmatrix}\xrightarrow{d}N_{|\mathcal{S}|+2}
\left[\begin{pmatrix}\mathcal{I}_{\rho,\beta}\delta\\
0 \\
\Pi_{\mathcal{S}}\mathcal{I}_{\beta,\beta}\delta
\end{pmatrix},\mathcal{I}_{\mathcal{S}}\right],
\end{equation*}
\item  \begin{equation*}
n^{1/2}
\begin{pmatrix}
\hat{\rho}_{S}-\rho_{0}\\
\hat{\sigma}_{\mathcal{S}}^{2}-\sigma_{0}^{2}\\ \hat{\beta}_{S}
\end{pmatrix}\xrightarrow{d}N_{|\mathcal{S}|+2}
\left[\begin{pmatrix}m_{\mathcal{S},\rho}\delta\\
m_{\mathcal{S},\sigma^{2}}\delta \\
m_{\mathcal{S},\beta_{\mathcal{S}}}\delta
\end{pmatrix},\mathcal{I}_{\mathcal{S}}^{-1}\right],
\end{equation*}
\end{enumerate}
where $m_{\mathcal{S},\rho}=\mathcal{I}_{\mathcal{S}}^{\rho,\rho}\mathcal{I}_{\rho,\beta}+\mathcal{I}_{\mathcal{S}}^{\beta,\rho}\Pi_{\mathcal{S}}\mathcal{I}_{\beta,\beta}$, $m_{\mathcal{S},\sigma^{2}}=\mathcal{I}_{\mathcal{S}}^{\sigma^{2},\rho}\mathcal{I}_{\rho,\beta}$, and $m_{\mathcal{S},\beta_{\mathcal{S}}}=\mathcal{I}_{\mathcal{S}}^{\beta,\rho}\mathcal{I}_{\rho,\beta}+\mathcal{I}_{\mathcal{S}}^{\beta,\beta}\Pi_{\mathcal{S}}\mathcal{I}_{\beta,\beta}$  are the matrices of orders $1\times p$, $1\times p$, and $|\mathcal{S}|\times p$ respectively. 
\end{theorem}
\section{The spatial FIC} \label{sec3}
\subsection{The asymptotic mean squared error of the estimated focus}
 Let $\mu_{W}^{*}=\mu_{W}(\rho,\sigma^2,\beta_{n})$ denote the least false focus function evaluated under the local misspecification \eqref{2.6}. Let
$J_{W,\mathcal{S}}=\frac{\partial\mu_{W}}{\partial\theta_{\mathcal{S}}}$
denote $k\times (|\mathcal{S}|+2)$ Jacobian matrix of $\mu_{W}$ with respect to $\theta_{\mathcal{S}}$. 
The following theorem provides the asymptotic normality of the focus function estimated under a locally misspecified subset model $\mathcal{S}$. 
\begin{theorem}\label{Th:2}
Under the setup of Theorem \ref{Th:1}, it holds that
\begin{equation*}
    n^{1/2}\left(\hat\mu_{W,\mathcal{S}}
    -\mu_{W}^{*}\right)\xrightarrow{d}N_{k}\left(b_{W,\mathcal{S}}\delta,V_{W,\mathcal{S}}\right)
\end{equation*}
as $n\to\infty$, where 
$\hat\mu_{W,\mathcal{S}}$ is the estimated focus function under a submodel $\mathcal{S}$, 
\begin{equation*}
m_{\mathcal{S}}=\left[m_{\mathcal{S},\rho}, m_{\mathcal{S},\sigma^{2}},m_{\mathcal{S},\beta_{\mathcal{S}}}\right]^{\top}, 
\end{equation*}
is an $(|\mathcal{S}|+2)\times p$ matrix, \begin{equation}
b_{W,\mathcal{S}}=J_{W,\mathcal{S}}m_{\mathcal{S}} \label{bias}
\end{equation}  and \begin{equation}
V_{W,\mathcal{S}}=J_{W,\mathcal{S}}\mathcal{I}_{\mathcal{S}}^{-1}J_{W,\mathcal{S}}^{\top} \label{variance}.
\end{equation}
\end{theorem}
Due to Equations (\ref{bias}) and (\ref{variance}), the asymptotic mean squared error (AMSE) of $\hat\mu_{W,\mathcal{S}}$ is hence given as 
\begin{equation}
\text{AMSE}(\hat\mu_{W,\mathcal{S}})=\text{Tr}\left[b_{W,\mathcal{S}}\delta\delta^{\top}b_{W,\mathcal{S}}^{\top}+V_{W,\mathcal{S}}\right]. \label{AMSE} 
\end{equation}

\subsection{Estimation of local misspecification parameter}
In order to estimate the squared bias term in Equation \eqref{AMSE}, the term $\Delta=\delta\delta^{\top}$ needs to be estimated. Let $\hat{\beta}_{\text{wide}}$ denote the estimator of $\beta$ under the wide model with  $\mathcal{S}=\{1,2,...,p\}$. The following theorem leads to an estimator of $\Delta$. 
\begin{theorem}\label{Th:3}
Under the setup of Theorem \ref{Th:1}, it holds that
\begin{equation*}
n^{1/2}\hat{\beta}_{\text{wide}}\xrightarrow{d}N\left(\delta,\mathcal{I}^{\beta,\beta}\right)
\end{equation*}
as $n\to\infty$. 
\end{theorem}
Define $D_{n}=n^{1/2}\hat{\beta}_{\text{wide}}$. Due to Theorem \ref{Th:3}, it is straightforward to see that the asymptotic mean of $D_{n}$ is $\delta$. Hence $D_{n}D_{n}^{\top}$ works as a plug-in estimator of $\Delta$. 
\subsection{The algorithm for FIC computation}\label{algorithm}
\begin{enumerate}
\item  Estimate the parameters $\rho, \sigma^2 ,\beta$ under the full model by maximizing the likelihood \eqref{likelihood}.
 \item     Estimate the Hessian $\mathcal{I}(\theta)$ and $\mathcal{I}(\theta)^{-1}$ and partition it according to $(\rho, \sigma^2 ,\beta)$ in that order. 
 \item         Estimate the term $\Delta$ using its plug-in estimator $D_{n}D_{n}^{\top}$ with $D_{n}$ as given by Theorem \ref{Th:3}.
 \item         For each submodel $\mathcal{S}$, carry out the following.
        \begin{enumerate}
  \item         Calculate the projection matrix $\Pi_{\mathcal{S}}$.
   \item        Partition both the Hessian and its inverse into the block components as given in Section \ref{sec2.3}.
   \item           Define the focus function and calculate its Jacobian $J_{W,\mathcal{S}}$. 
   \item           Calculate the matrix $m_{\mathcal{S}}$ as in Theorem \ref{Th:2}. 
   \item           Estimate the bias $b_{W,\mathcal{S}}$ and the variance $V_{W,\mathcal{S}}$ as given by equations \eqref{bias} and \eqref{variance} respectively. 
   \item           Calculate the FIC value as the estimate of the AMSE given by Equation \eqref{AMSE}. 
            
        \end{enumerate}
     \item     Rank the models in ascending order of the FIC values. 
        \end{enumerate}
        \section{The spatial average FIC (sAFIC)} \label{sec4}
        \subsection{Background}
        The linear predictor at a fixed covariate level as given in Equation \eqref{focus1} is one of the most common and intuitive choices of the focus functions for variable selection. However, the main drawback of this approach is its inability to incorporate varying covariate levels while evaluating the models based on the FIC. Covariate levels at different regions in principle may lead to different model choices. In order to select a single most suited model for the entire spatial data across a range of regions, it is necessary to devise a suitable summary of the AMSE based on the aggregation of the varying covariate levels. \citet{claeskens2003focused, claeskens2008model}, proposed the averaged version of the FIC using the integrated mean squared error of the estimated focus with respect to some relevant distribution of the covariate levels. The authors termed it as the average FIC. In this section, we propose the spatial version of this average FIC for the SLM model under consideration. We term this as the sAFIC criterion.  
        \subsection{The limiting distribution of estimated linear predictor}
        In Theorem \ref{Th:2}, we provide the limiting distribution of the estimator of a generic vector-valued focus function. In this section, we provide its simplified form when the focus is a linear predictor. This simplified result is instrumental in obtaining the integrated mean squared error of the linear predictor leading to the desired sAFIC expressions. 
        \par Observe that in order to obtain the sAFIC expression, we need to obtain the asymptotic distribution of $n^{1/2}\left(\hat{\mu}_{W,i,\mathcal{S}}-\mu_{W,i}^{*}\right)$ where $\hat{\mu}_{W,i,\mathcal{S}}$ denotes the linear predictor corresponding to region $i$ while $\mu_{W,i}^{*}$ is the ``least false" value of the focus function evaluated under the local misspecification. Consider the partitioning of the Fisher information matrix as per $(\rho,\beta)$ as follows. 
        \begin{equation*}
\begin{pmatrix}
\mathcal{I}_{\rho,\rho} & \mathcal{I}_{\rho,\beta}\Pi_{\mathcal{S}}^{\top} \\
\Pi_{\mathcal{S}}\mathcal{I}_{\beta,\rho}  & \Pi_{\mathcal{S}}\mathcal{I}_{\beta,\beta} \Pi_{\mathcal{S}}^{\top} 
\end{pmatrix}
\end{equation*} while its inverse is partitioned similarly.  
The following theorem provides the limiting distribution of the estimated linear predictor. In what follows, $U_{n,\rho}$ and $U_{n,\beta}$ denote the gradients of the log-likelihood of the SLM Model \eqref{model} while $U_{\rho}$ and $U_{\beta}$ denote the limiting random variables corresponding to $n^{-1/2}U_{n,\rho}$ and $n^{-1/2}U_{n,\beta}$ respectively. The matrix $Q=\left(\mathcal{I}_{\beta,\beta}-\mathcal{I}_{\beta,\rho}\mathcal{I}_{\rho,\rho}^{-1}\mathcal{I}_{\rho,\beta}\right)^{-1}$ denotes the component corresponding to $I_{\beta,\beta}$ in the inverse Fisher information while $Q_{\mathcal{S}}=\left(\pi_{\mathcal{S}}Q^{-1}\pi_{\mathcal{S}}^{\top}\right)^{-1}$ denotes the submatrix of $Q$ corresponding to a subset model $\mathcal{S}$. 
        \begin{theorem}\label{Th:4}
        Under the setup of Theorem \ref{Th:1}, the following  holds true as $n\to\infty$. 
       
\begin{equation*}n^{1/2}\left(\hat{\mu}_{W,i,\mathcal{S}}-\mu_{W,i}^{*}\right)\xrightarrow{d}\Lambda_{W,i,\mathcal{S}}=\Lambda_{W,i}+\omega_{W,i}^{\top}\left(\delta-G_{\mathcal{S}}C\right), \end{equation*}
    where
     \begin{equation*}
\Lambda_{W,i}=\mathcal{I}_{\rho,\rho}^{-1}U_{\rho}\sum_{j=1}^{n}w_{ij}y_{j},
     \end{equation*}
     \begin{equation*}
    \omega_{W,i}=\mathcal{I}_{\beta,\rho}\mathcal{I}_{\rho,\rho}^{-1}\sum_{j=1}^{n}w_{ij}y_{j}-x_{i},
     \end{equation*}
     \begin{equation*}
G_{\mathcal{S}}=\pi_{\mathcal{S}}^{\top}Q_{{\mathcal{S}}}\pi_{\mathcal{S}}Q^{-1}
     \end{equation*}
     and 
     \begin{equation*}
         C=\delta+Q\left(U_{\beta}-\mathcal{I}_{\beta,\rho}\mathcal{I}_{\rho,\rho}^{-1}U_{\rho}\right).
         \end{equation*}
Moreover, the random variables $\Lambda_{W,i}$ and $\omega_{W,i}^{\top}\left(\delta-G_{\mathcal{S}}A\right)$ are independent of each other for any fixed $i$. 
      \end{theorem}
      Due to Theorem \ref{Th:4}, it can be seen that the AMSE of the estimated linear predictor estimated under a locally misspecified subset model $\mathcal{S}$ is given as
        \begin{equation}\label{AMSE1}
           R_{W,i,\mathcal{S}}=\text{AMSE}(\hat{\mu}_{W,i,\mathcal{S}})= \omega_{W,i}^{\top}\left(I_{p}-G_{\mathcal{S}}\right)\delta\delta^{\top}\left(I_{p}-G_{\mathcal{S}}\right)^{\top}\omega_{W,i}+\left(\frac{\partial\mu_{W,i}}{\partial\rho}\right)^{2}\mathcal{I}_{\rho,\rho}^{-1}+\omega_{W,i}^{\top}G_{\mathcal{S}}QG_{\mathcal{S}}^{\top}\omega_{W,i}.
        \end{equation}
    
        \subsection{sAFIC expressions}
In order to develop the averaged version of the FIC, we need to consider the integrated version of the limiting risk $R_{W,i,\mathcal{S}}$ as given by Equation \eqref{AMSE1} taking into account the distribution of the covariate levels. For this purpose, we consider the rows of the design matrix $X$ to be $n$ realizations of a random vector $Z$ taking values in $\mathbb{R}^{p}$. Let $F(z)$ denote the distribution function of $Z$ evaluated at $z\in\mathbb{R}^{p}$. We assume throughout these discussions that $Z$ is square-integrable with respect to $F$. Let $R_{W,\mathcal{S}}(z)$ denote a random variable whose realizations are $\{R_{W,i,\mathcal{S}},i=1,2,\cdots,n\}$. 
Suppose $\psi$ is a  bounded, non-negative function on $R^{p}$ that is integrable with respect to $F$. Without loss of generality, assume that  $\int_{\mathbb{R}^{p}}\psi(z) dF(z)=1$. Such a structure for $\psi$ ensures that it defines a valid weighting scheme that can be used to obtain the weighted distribution of the covariates. In practice, weights may be chosen based on the covariate levels for a cluster of regions in the given data. 
\par With respect to the weight function $\psi$, define the weighted distribution function as
\begin{equation*}
F_{\psi}(z)=\int_{z^{*}\in(-\infty,z]^{p}}\psi(z^{*})dF(z^{*}). 
\end{equation*}
The integrated version of the risk $R_{W,\mathcal{S}}(z)$ can now be given as 
\begin{equation*}
\eta_{W,\psi,\mathcal{S}}=\int_{\mathbb{R}^{p}}R_{W,\mathcal{S}}(z)dF_{\psi}(z)
\end{equation*}
The above function is interpreted as the weighted mean of the risk $R_{W,\mathcal{S}}$ where the weight function is $\psi(z)$. 
\par The following theorem provides the explicit formula for $\eta_{W,\psi,\mathcal{S}}$ that is instrumental in obtaining the compact averaged FIC formula. In what follows, the term $\mu_{W}(z)$ denotes the real-valued random focus function (linear predictor) whose realizations are $\{\mu_{W,i},i=1,2,\cdots,n\}$. Similarly $\omega_{W}(z)$ is a random vector taking values in $\mathbb{R}^{p}$ whose realizations are $\{\omega_{W,i},i=1,2,\cdots,n\}$ where $\omega_{W,i}$ is as given by Theorem \ref{Th:4}. 
\begin{theorem}\label{Th:5}
For a given spatial adjacency matrix $W$ and the weight function $\psi$, let $H_{W,\psi}$ denote $(p+1)\times (p+1)$ generalized information matrix given  as 
\begin{equation*}
H_{W,\psi}=\int_{\mathbb{R}^{p}}\begin{pmatrix}
\frac{\partial\mu_{W}(z)}{\partial\rho}\\
\frac{\partial\mu_{W}(z)}{\partial\beta}
\end{pmatrix}
\begin{pmatrix}
\frac{\partial\mu_{W}(z)}{\partial\rho}\\
\frac{\partial\mu_{W}(z)}{\partial\beta}
\end{pmatrix}^{\top}dF_{\psi}(z)
=\begin{pmatrix}
H_{W,\psi,\rho,\rho}   & H_{W,\psi,\rho,\beta} \\
H_{W,\psi,\beta,\rho}   & H_{W,\psi,\beta,\beta}
\end{pmatrix}.
\end{equation*}
Define $p\times p$ matrix $K_{W,\psi}$ as 
\begin{equation*}
K_{W,\psi}=\int_{\mathbb{R}^{p}} \omega_{W}(z)\omega_{W}(z)^{\top}dF_{\psi}(z)=\mathcal{I}_{\beta,\rho}\mathcal{I}_{\rho,\rho}^{-1}H_{W,\psi,\rho,\rho}\mathcal{I}_{\rho,\rho}^{-1}\mathcal{I}_{\rho,\beta}-2\mathcal{I}_{\beta,\rho}\mathcal{I}_{\rho,\rho}^{-1}H_{W,\psi,\rho,\beta}+H_{W,\psi,\beta,\beta}.
\end{equation*}
The weighted mean $\eta_{W,\psi,\mathcal{S}}$ is given as 
\begin{equation*}
\eta_{W,\psi,\mathcal{S}}=\text{Tr}\left[\left(I_{p}-G_{\mathcal{S}}\right)\delta\delta^{\top}\left(I_{p}-G_{\mathcal{S}}\right)^{\top}K_{W,\psi}\right]+{\text{Tr}\left(G_{\mathcal{S}}QG_{\mathcal{S}}^{\top}K_{W,\psi}\right)}+\int \left(\frac{\partial\mu_{W}(z)}{\partial\rho}\right)^{2}\mathcal{I}_{\rho,\rho}^{-1}\psi(z)dF_{\psi}(z).
\end{equation*}
\end{theorem}
\noindent In order to devise the sAFIC formula, we need the empirical version of $\eta_{W,\psi,\mathcal{S}}$. Based on a random sample $\{z_{1},\cdots,z_{n}\}$, the empirical weights are chosen such that $\psi(z_{i})\geq 0$ for every $i\in\{1,2,\cdots,n\}$ and $
\sum_{i=1}^n \psi(z_{i})=1. 
$
Let $H_{n,W,\psi}$ denote the empirical version of $H_{W,\psi}$.  It can be seen that 
\begin{equation*}
\begin{split}
H_{n,W,\psi}&=\sum_{i=1}^{n}\psi(z_{i})\begin{pmatrix}
\frac{\partial\mu_{W}(z_{i})}{\partial\rho}\\
\frac{\partial\mu_{W}(z_{i})}{\partial\beta}
\end{pmatrix}
\begin{pmatrix}
\frac{\partial\mu_{W}(z_{i})}{\partial\rho}\\
\frac{\partial\mu_{W}(z_{i})}{\partial\beta}
\end{pmatrix}^{\top}
\\
&=
\begin{pmatrix}
Y^{\top}W^{\top}\Psi WY   & Y^{\top}W^{\top}\Psi X \\
X^{\top}\Psi WY   & X^{\top}\Psi X
\end{pmatrix}
\end{split}
\end{equation*}
where $\Psi=\text{diag}\left(\psi(z_{1}),\cdots,\psi(z_{n})\right)$ is $n\times n$ diagonal weight matrix. Let $K_{n,W,\psi}$ denote the empirical version of $K_{W,\psi}$ that is defined in the same spirit. The sAFIC with respect to the spatial connectivity matrix $W$ and the weighted matrix $\Psi$ is given as  
\begin{equation}
\text{sAFIC}(\Psi,W,\mu_{W,i,\mathcal{S}})=\text{Tr}\left[\left(I_{p}-G_{n,\mathcal{S}}\right)D_{n}D_{n}^{\top}\left(I_{p}-G_{n,\mathcal{S}}\right)^{\top}K_{n,W}\right]+{\text{Tr}\left(G_{n,\mathcal{S}}Q_{n}G_{n,\mathcal{S}}^{\top}K_{n,W}\right)}, \label{sAFIC_exp}
\end{equation}
where $G_{n,\mathcal{S}}$ and $Q_{n}$ are the sample versions of the matrices $G_{\mathcal{S}}$ and $Q$ as defined in Theorem \ref{Th:4}.
\begin{remark}
Note that the last term $\int \left(\frac{\partial\mu_{W}(z)}{\partial\rho}\right)^{2}\mathcal{I}_{\rho,\rho}^{-1}dF_{\psi}(z)$ appearing in $\eta_{W,\mathcal{S}}$ is common to all subset models under consideration and, therefore, can be ignored while comparing the IAMSE of different models. Hence, this term does not appear in the sAFIC formula as given by Equation \eqref{sAFIC_exp}. 
\end{remark}
\subsection{Choice of the weight functions}
\begin{enumerate}
   \item   \textbf{Uniform weights}~\\In this setup, we take 
     \begin{equation}
\psi(z_{i})=1/n  \ \ , i=1,2,\cdots n. \label{safic1}
  \end{equation}
With this choice, the weighted mean as given by Theorem \ref{Th:5} reduces to the usual arithmetic mean and the resulting sAFIC is simply based on the empirical distribution function $F_{n}(z)$. 
      \item \textbf{Kernel based weights}~\\
   Another choice of the weight functions is 
  \begin{equation}
      \psi(z_{i})=\text{Ker}\left(\frac{||z-z_{i}||}{h}\right), \label{safic2}
  \end{equation}
  where $\text{Ker}$ denotes the Kernel function used for the nonparametric density estimation (\cite{hollander2013nonparametric}, Chapter 12), $h$ denotes the bandwidth and $z$ is a fixed covariate level. Such a choice of weights takes into account the local influence of each covariate level $z_{i}$ on the predetermined level $z$. The choice of $h$ determines the scale of smoothing in the estimation of the density of the covariates thereby allowing the closer examination of the bias-variance trade-off in the density estimation. In principle, the predetermined covariate level $z$ can correspond to the covariate levels for the regions of interest such as the hotspots in the spatial data. In this way, the impact of the hotspot regions is direcly taken into account while building the models. One of the most widely used  kernels is the Gaussian kernel given as 
   $$\text{Ker}(||u||)=(2\pi)^{-p/2}\exp\left(-\frac{1}{2}||u||^2\right),u\in\mathbb{R}^{p}.$$ 
   \end{enumerate}   
\begin{remark}
The sAFIC is comparable with the traditional AIC-like information criteria which involve the penalty term for the model complexity. Note that the term $\text{Tr}\left(G_{n,\mathcal{S}}Q_{n}G_{n,\mathcal{S}}^{\top}K_{n}\right)$ acts as a penalty for overfitting like the one in the AIC-like criteria while the term $\text{Tr}\left[\left(I_{p}-G_{n,\mathcal{S}}\right)D_{n}D_{n}^{\top}\left(I_{p}-G_{n,\mathcal{S}}\right)^{\top}K_{n}\right]$ acts as a model deviance. More details on the comparisons of the AIC and the averaged FIC can be found in \cite{pandhare2020focussed}.
\end{remark}

\section{Simulations} \label{sec5}
    In our simulation study, we generate data using the SLM as given by Equation (\ref{model}). The error terms are assumed to be iid standard normal variables. The number of spatial units is taken to be $n=75$, and the parameter for spatial dependence is set to $\rho=0.5$. We use a first-order spatial adjacency matrix $W$, representing the Lag-1 neighborhood structure. The true regression coefficient vector consists of ${\beta}=(0,0.2,0.2,0,0)^{\top}$. The narrow model is a pure SLM with no exogeneous covariates and the wide model has 5 covariates. Hence we now have a total of $2^{5}=32$ choices of submodels  as listed in the Table \ref{tab:models}. In the following table under each submodel we indicate the variables excluded and included using $0$ and $1$ respectively.
         
\begin{table}[h]
    \centering
    \begin{tabular}{lccccc}
        \toprule
        \textbf{Submodel} & \textbf{Var1} & \textbf{Var2} & \textbf{Var3} & \textbf{Var4} & \textbf{Var5} \\
        \midrule
        \textbf{Narrow Model (S1)} & 0 & 0 & 0 & 0 & 0 \\
        $S_2$  & 1 & 0 & 0 & 0 & 0  \\
        $S_3$  & 0 & 1 & 0 & 0 & 0  \\
        \vdots & \vdots & \vdots & \vdots & \vdots & \vdots  \\
        $S_7$ & 0 & 1 & 1 & 0 & 0  \\
        \vdots & \vdots & \vdots & \vdots & \vdots & \vdots  \\
        \textbf{Wide Model} ($S_{32}$) & 1 & 1 & 1 & 1 & 1  \\
        \bottomrule
    \end{tabular}
    \caption{Selective submodels ranging from the Narrow Model ($S_1$) to the Wide Model ($S_{32}$).}
    \label{tab:models}
\end{table}
\noindent The FIC scores under each focus function for all possible submodels, are summarized in Table \ref{tab:criteria_ranking}. Top $5$ models along with their frequency of selection in $100$ simulation runs are reported.
\begin{table}[H]
    \centering
    \caption{Ranking of different criteria based on frequency}
    \label{tab:criteria_ranking}
    \renewcommand{\arraystretch}{1.2}
    \small
    \begin{tabular}{l c c c c c}
        \toprule
        \textbf{Criteria} & \textbf{Rank 1} & \textbf{Rank 2} & \textbf{Rank 3} & \textbf{Rank 4} & \textbf{Rank 5} \\ 
        \midrule
        \multirow{2}{*}{FIC1 \eqref{focus1}}  & $S_{17} (24)$  & $S_{3} (13)$  & $S_{9} (12)$  & $S_{2} (11)$ & $S_{5} (6)$ \\  
                               & $(X_5)$ & $(X_2)$ & $(X_4)$ & $(X_1)$ & $(X_3)$ \\ 
        \midrule
        \multirow{2}{*}{FIC2 \eqref{focus3}}  & $S_{9} (16)$  & $S_{2} (13)$  & $S_{17} (13)$  & $S_{25} (7)$ & $S_{19} (6)$ \\  
                               & $(X_4)$ & $(X_1)$ & $(X_5)$ & $(X_4,X_5)$ & $(X_2,X_5)$ \\ 
        \midrule
        \multirow{2}{*}{VFIC1\eqref{focus2}}  & $S_{17} (35)$  & $S_{9} (25)$  & $S_{2} (24)$  & $S_{5} (10)$ & $S_{3} (8)$ \\  
                                & $(X_5)$ & $(X_4)$ & $(X_1)$ & $(X_3)$ & $(X_2)$ \\  
        \midrule
        \multirow{2}{*}{VFIC2 \eqref{focus4}}  & $S_{2} (25)$  & $S_{9} (22)$  & $S_{17} (20)$  & $S_{5} (19)$ & $S_{3} (14)$ \\  
                                & $(X_1)$ & $(X_4)$ & $(X_5)$ & $(X_3)$ & $(X_2)$ \\  
        \midrule
        \multirow{2}{*}{sAFIC1 \eqref{safic1}}  & $S_{32}(75)$  & $S_{16} (7)$  & $S_{24} (5)$  & $S_{31} (5)$ & $S_{28} (3)$ \\  
                                 & $(X_1, X_2, X_3, X_4, X_5)$ & $(X_1, X_2, X_3, X_4)$ & $(X_1,X_2, X_3, X_5)$ & $(X_2,X_3, X_4, X_5)$ & $(X_1,X_2, X_4, X_5)$ \\  
        \midrule
        \multirow{2}{*}{sAFIC2 \eqref{safic2}}  & $S_{32}(75)$  & $S_{16} (7)$  & $S_{24} (6)$  & $S_{31} (5)$ & $S_{28} (3)$ \\  
                                 & $(X_1, X_2, X_3, X_4, X_5)$ & $(X_1, X_2, X_3, X_4)$ & $(X_1,X_2, X_3, X_5)$ & $(X_2,X_3, X_4, X_5)$ & $(X_1,X_3, X_4, X_5)$ \\  
        \midrule
        \multirow{2}{*}{AIC}  & $S_{32} (67)$  & $S_{24} (10)$  & $S_{16} (9)$  & $S_{31} (6)$ & $S_{30} (3)$ \\  
                              & $(X_1, X_2, X_3, X_4, X_5)$ & $(X_1, X_2, X_3, X_5)$ & $(X_1,X_2, X_3, X_4)$ & $(X_2,X_3, X_4, X_5)$ & $(X_1,X_2, X_4, X_5)$ \\  
        \bottomrule
    \end{tabular}
\end{table}
  The FIC values computed from focus functions \eqref{focus1}–\eqref{focus4} lead to the selection of different submodels. Variable selection using the sAFIC with weight functions as in \eqref{safic1} and \eqref{safic2} is comparable to that of the AIC. Both these criteria are seen to prefer models closer to the wide model. While simulations indicate that varying the weights in sAFIC has little impact on overall performance, we highlight their differing results in the application section to follow. However it is crucial to note that the sAFIC shares the advantages of the focussed model selection as well as the traditional goodness of fit based information criteria like AIC.
\section{Application to real data} \label{sec6}
We demonstrate an application of the proposed focused variable selection using the Boston Housing Dataset. This dataset consists 506 observations from Boston’s census tracts, capturing demographic, economic, environmental, and geographic factors. The response variable is the median housing value (MEDV), with covariates as crime rate (CRIM), air pollution (NOX), land use (ZN, INDUS), accessibility (DIS, RAD), and housing characteristics (RM, AGE). The dataset also consists of socioeconomic indicators such as lower-status population (LSTAT), tax, pupil-teacher ratio (PTRATIO), and Black population proportion (B) offer insights into neighborhood conditions. Geographic location is specified by latitude (LAT) and longitude (LON). For more details, see the dataset documentation: \url{https://www.rdocumentation.org/packages/MASS/versions/7.3-64/topics/Boston}.
\par
\begin{figure}[H]
    \centering
\includegraphics[width=\linewidth]{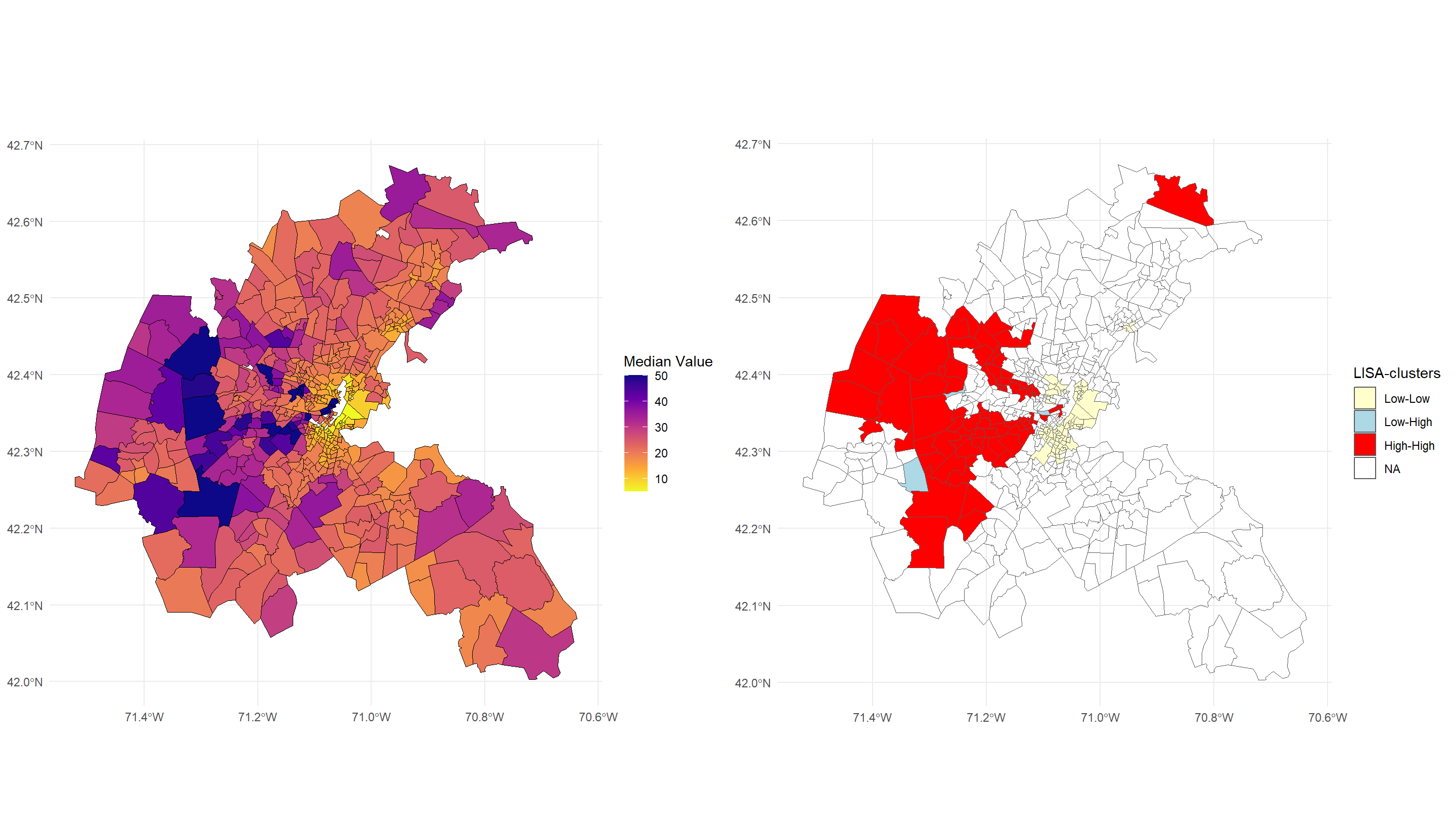}
    \caption{Spatial distribution of median housing values across $506$ tracts in Boston (left); Significant spatial clusters identified from local spatial autocorrelation analysis (right).}
    \label{fig:1}
\end{figure}

 The right panel of Figure[\ref{fig:1}] shows the spatial distribution of our response variable across 506 tracts in Boston. The figure reveals clear spatial patterns in the distribution of housing values. Furthermore, the presence of significant positive spatial autocorrelation is confirmed by the Moran's $I$ test. The significantly positive Moran’s I statistic indicates strong spatial autocorrelation in housing prices, meaning that highly priced houses tend to cluster near other highly priced houses. By means of the local spatial autocorrelation analysis, we identified these spatial clusters with the results shown in the left panel of Figure[\ref{fig:1}].
 For the spatial autocorrelation analysis, a Lag-1 adjacency structure was used to define spatial neighborhood, ensuring that each unit is analyzed in relation to its immediate neighbors. This spatial dependence underscores the importance of incorporating spatial models in the analysis, as traditional regression models may not take into account the underlying spatial structure.
\par The above exploratory analysis reveals significant spatial autocorrelation in the data hence we fit the SLM given by \eqref{model} to the Boston data. 
It has been observed that CRIM, NOX, TAX, DIS, and PTRATIO have significant negative effects on the housing prices. Conversely variables indicating housing conditions and accessibility, such as ZN, RM, and RAD exihibit significant positive effects. 
\begin{table}[h]
    \centering
    \renewcommand{\arraystretch}{1.2} 
    \resizebox{\textwidth}{!}{ 
        \begin{tabular}{l*{12}{c}} 
            \toprule
            Criteria & CRIM & ZN & INDUS & NOX & RM & AGE & DIS & RAD & TAX & PTRATIO & B & LSTAT \\
            \midrule
            FIC$_H$ \eqref{focus1} & \ding{51} & \ding{51} &  & \ding{51} & &  &\ding{51}  &  &  &  &  &  \\
            FIC$_L$ \eqref{focus1} &  &\ding{51}  &\ding{51}  &  &  &  &\ding{51}  &  &  & \ding{51} &  &  \\
            FIC$_2$ \eqref{focus3} &  &\ding{51}  & \ding{51} & \ding{51} &  &  & &  &  & \ding{51} & \ding{51}  &  \\
            VFIC \eqref{focus2} &  &  &  &\ding{51}  &  & \ding{51} &  & \ding{51} &  &  & \ding{51} &  \\
            VFIC \eqref{focus4} &  &  &  & \ding{51}  &  \ding{51} &  &  &  & \ding{51} & \ding{51} & \ding{51} & \ding{51}  \\
            sAFIC$_{H-H}$ \eqref{safic1} & &\ding{51}  & & \ding{51}  &\ding{51}  &  &\ding{51}  & &\ding{51}  &\ding{51}  &\ding{51}  & \ding{51} \\
            sAFIC$_{L-L}$ \eqref{safic1} & \ding{51} & \ding{51}  &  & \ding{51} & \ding{51} &  & \ding{51} & \ding{51} & &\ding{51}  &\ding{51}  &\ding{51} \\
             sAFIC \eqref{safic2} &  &  & & \ding{51}  & \ding{51}  & \ding{51} &\ding{51}   &\ding{51}   &  &  \ding{51} & \ding{51}  & \ding{51}  \\
            AIC & \ding{51} & \ding{51} &  & \ding{51} & \ding{51} &  & \ding{51} & \ding{51} & \ding{51} & \ding{51} & \ding{51} & \ding{51} \\
            \bottomrule
        \end{tabular}
    }
    \caption{Variables selected across different focused criteria based on the top 100 submodels ranked by FIC score.}
    \label{tab:1}
\end{table}
\par However, given the heterogeneity among the tracts not all these variables may be important across all spatial units. We demonstrate this argument using our proposed methodology, which results in a contextual variable selection.
\begin{figure}[H]
    \centering
\includegraphics[width=0.5\linewidth]{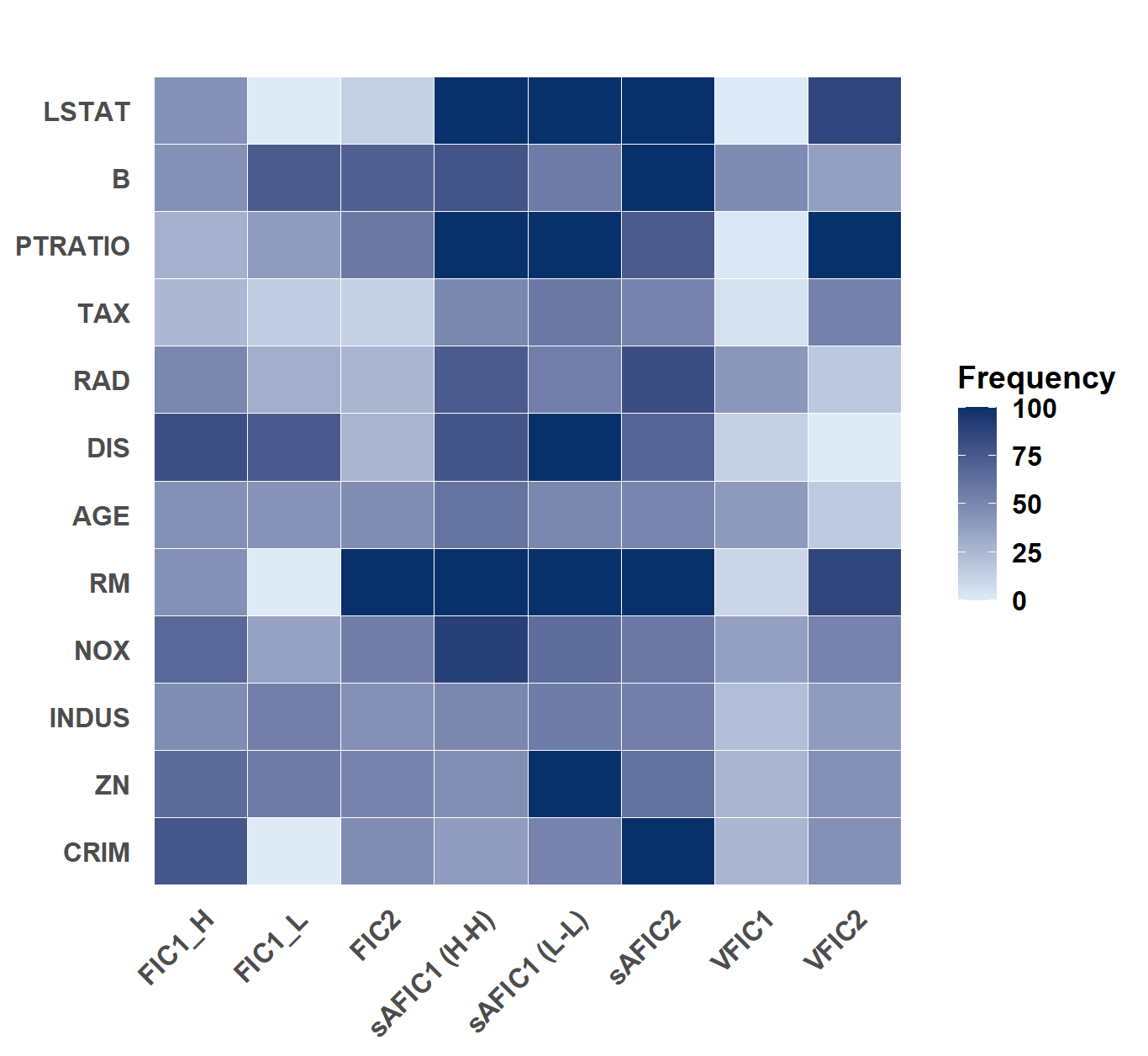}
    \caption{Frequency of variables among the top $100$ submodels ranked based on FIC score across different focused criteria.}
    \label{fig:2}
\end{figure}
\par Note that our first focus function defined in \eqref{focus1}, allows flexible variable selection driven by local effects in the data. In Table \ref{tab:1} we present the results of variable selection based on the spatial units with high (FIC$_H$) and low (FIC$_L$) median housing value using focus function \eqref{focus1}.
Moreover, the focus functions \eqref{focus3}, \eqref{focus2}, and \eqref{focus4} yield notably different sets of selected variables. Instead of a single covariate level we might want to look at the variables selections based on cluster of regions with High/Low response values. This clustering is addressed through the our spatially averaged FIC (sAFIC) using equal weights \eqref{safic1}, as well as a Gaussian kernel-based weighting scheme \eqref{safic2}. For completeness, we also report variables selected via AIC for comparison. Figure \ref{fig:2} illustrates the distribution of variable selection frequency across the top $100$ submodels ranked by FIC score. Notably, CRIM, ZN, NOX, and DIS are frequently selected under the high MEDV regions, whereas ZN, INDUS, DIS, and the proportion of Black population are more prominent in the low MEDV regions. Moreover, the focus functions \eqref{focus2}, \eqref{focus3}, and \eqref{focus4} yield notably different sets of selected variables.
\section{Concluding remarks and future directions} \label{sec7}
In this paper, we have devised the FIC-based variable selection methodology for spatial lag models. Given the spatial contiguity structure, the proposed method aims to identify a subset of exogenous predictors that minimizes the risk in estimating the user-specified focus function under potential model misspecification. The FIC methodology leads to contextually useful and interpretable models by evaluating the bias-variance trade-off in the estimation under model misspecification. Different choices of the focus functions, in principle, lead to different model choices. We suggest practically useful choices of the focus functions that account for the impact of the covariates on the average response, the variability of the estimators and spatial spillover effects. The systematic asymptotic analysis is carried out leading to the desired FIC expressions. We have also developed the weighted averaged version of the FIC where the weights are chosen based on the distribution of the covariate levels. Such an averaged version of the FIC leads to models that perform well for a cluster of spatial regions in the data rather than merely for the predetermined regions or certain hot-spots. The simulation experiments demonstrate the satisfactory performance of the proposed methodology. The FIC is seen to select models with lower realized mean squared error of the estimated focus.
\par The methodology proposed in this paper can be implemented for a variety of spatial regression models. Below we sketch the FIC formulation for two most commonly used spatial models, namely, spatial lag and error models and spatial regression models. 
\subsection{Spatial lag and error model}
The proposed methodology can be readily extended to address the variable selection problem when the innovations in the SLM are no longer independently distributed but are rather spatially correlated. The spatial lag and error models combine both the spatially lag response and spatially lagged error thereby treating the response to be spatially autoregressive moving average process.
\par Consider a spatial lag and error model given as 
\begin{equation*}
\begin{split}
Y&=\rho WY+X \beta + \epsilon\\
&\epsilon=\lambda M \epsilon+\nu
\end{split}
\end{equation*}
where $M$ denotes the spatial adjacency matrix corresponding to the errors $\epsilon$, $\lambda$ denotes the corresponding spatial error coefficient while $\nu=\left(\nu_{1},\cdots,\nu_{n}\right)^{\top}$ denotes the vector of iid errors. It is of interest to select the columns of $X$ using the FIC. Without loss of generality, the narrow model in this case can be considered to be purely spatial lag model with $\lambda=0$ and $\beta=0$. Hence, the protected parameters common to all the models are $(\rho,\sigma^{2})$, while the additional parameters subject to inclusion or exclusion are $(\lambda,\beta)$. As in Section \ref{sec2.3}, the $(|\mathcal{S}|+3)\times (|\mathcal{S}|+3)$ Fisher information can be partitioned as 
\begin{equation}
\mathcal{I}_{\mathcal{S}}=\begin{pmatrix}
\mathcal{I}_{\rho,\rho} & \mathcal{I}_{\rho,\sigma^{2}} &\mathcal{I}_{\rho,\lambda} & \mathcal{I}_{\rho,\beta}\Pi_{\mathcal{S}}^{\top} \\
\mathcal{I}_{\sigma^{2},\rho}& \mathcal{I}_{\sigma^{2},\sigma^{2}} &\mathcal{I}_{\sigma^{2},\lambda}  & O_{1\times |\mathcal{S}|}\\
\mathcal{I}_{\lambda,\rho}& \mathcal{I}_{\lambda,\sigma^{2}} &\mathcal{I}_{\lambda,\lambda}  & \mathcal{I}_{\lambda,\beta}\Pi_{\mathcal{S}}^{\top} \\
\Pi_{\mathcal{S}}\mathcal{I}_{\beta,\rho} & O_{|\mathcal{S}|\times 1} & \Pi_{\mathcal{S}}I_{\beta,\lambda}  &\Pi_{\mathcal{S}}I_{\beta,\beta} \Pi_{\mathcal{S}}^{\top} \\
\end{pmatrix}.
\end{equation}
The FIC can now be calculated following the algorithm in Section \ref{algorithm}. 
\subsection{Spatiotemporal regression models}
Consider the spatiotemporal regression with the maximal order of temporal dependence to be $q$
model given as 
\begin{equation*}
y_{it} = \sum_{k=1}^{q} \lambda_k y_{i,t-k} + \rho \sum_{j=1}^{N} w_{ij} y_{jt} + X_{it}^{\top} \beta + \varepsilon_{it}
\end{equation*}
\noindent where \( y_{it} \) denote the dependent variable for unit \( i \) at time \( t \), with \( \lambda_k \) representing the coefficients on the \( k \)th temporal lags. The spatial weights matrix \( W \) has elements \( w_{ij} \), and \( \rho \) denotes the spatial autoregressive parameter. The vector \( X_{it} \) is a \( p \times 1 \) set of covariates for unit \( i \) at time \( t \), with corresponding coefficient vector \( \beta \). The error terms \( \varepsilon_{it} \) are assumed to be independently and identically distributed as \( N(0, \sigma^2) \).
It is of interest to select the order of temporal dependence of the spatially varying response along with a subset of exogenous covariates $X_{i}$. The parameter vector $\theta$ is of the dimension $q+p+2$ with the elements $\theta=\left(\lambda,\rho,\beta,\sigma^2\right)$ where $\lambda=\left(\lambda_{1},\cdots,\lambda_{q}\right)^{\top}$. We assume that the smallest model is a purely spatiotemoral model with no covariates and of the temporal order $r\leq p$. Thus, the narrow model consists of $r+2$ parameters $(\alpha,\rho,\sigma^2)$ with $\alpha=(\lambda_{1},\cdots,\lambda_{r})^{\top}$. The additional $q-r+p$ parameters subject to selection are given by a vector $\left(\gamma,\beta\right)$ where $\gamma=\left(\lambda_{r+1},\cdots,\lambda_{q}\right)^{\top}$. Partition the $(q+|\mathcal{S}|+2)\times(q+|\mathcal{S}|+2) $ Fisher information matrix for a subset model $\mathcal{S}$ in the order $(\alpha,\rho,\sigma^{2},\lambda,\beta_{\mathcal{S}})$ as 
\begin{equation*}
\mathcal{I}_{\mathcal{S}}=\begin{pmatrix}
\mathcal{I}_{\alpha,\alpha} & \mathcal{I}_{\alpha,\rho} &O_{r\times 1} & \mathcal{I}_{\alpha,\gamma} & \mathcal{I}_{\alpha,\beta}\Pi_{\mathcal{S}}^{\top}\\
\mathcal{I}_{\rho,\alpha} & \mathcal{I}_{\rho,\rho} &\mathcal{I}_{\rho,\sigma^2} & \mathcal{I}_{\rho,\gamma} & \mathcal{I}_{\rho,\beta}\Pi_{\mathcal{S}}^{\top}\\
O_{1\times r}& \mathcal{I}_{\sigma^{2},\rho} &\mathcal{I}_{\sigma^{2},\sigma^2}  & O_{1\times (q-r)} & O_{1\times p}
\\
\mathcal{I}_{\gamma,\alpha} & \mathcal{I}_{\gamma,\rho} &O_{(q-r)\times 1} & \mathcal{I}_{\gamma,\gamma} & \mathcal{I}_{\gamma,\beta}\Pi_{\mathcal{S}}^{\top}\\
\Pi_{\mathcal{S}}\mathcal{I}_{\beta,\alpha} & \Pi_{\mathcal{S}}\mathcal{I}_{\beta,\rho} &O_{|\mathcal{S}|\times 1} & \Pi_{\mathcal{S}}\mathcal{I}_{\beta,\gamma} & \Pi_{\mathcal{S}}\mathcal{I}_{\beta,\beta}\Pi_{\mathcal{S}}^{\top}
\end{pmatrix}.
\end{equation*}
The FIC for each subset model can now be calculated following the routine in Section \ref{algorithm}. We leave further investigations for future considerations. 
\bibliographystyle{apalike}
\bibliography{references}
\clearpage
\appendix
\section{Appendix: Assumptions and related discussions}\label{App:A}
As in \cite{ord1975estimation}, we make the following assumptions which ensure the asymptotic efficiency of the MLE. In what follows, $\omega_{1}\cdots\omega_{n}$ denote the  eigenvalues of $W$ with $\omega_{min}$ and $\omega_{max}$ being the minimum and the maximum eigenvalue respectively.  
\begin{assumption}\label{A1}
All the diagonal elements of the spatial adjacency matrix $W$ are zeros. 
\end{assumption}

\begin{assumption}\label{A2}
All the eigenvalues of $W$ are real. 
\end{assumption}
\begin{assumption}\label{A3}
    The spatial autocorrelation parameter $\rho$ satisfies $-1/\omega_{min}<\rho<1/\omega_{max}$.
\end{assumption}
\begin{assumption}\label{A4}
    The design matrix $X$ is a full rank matrix for each $n$. Moreover,~\\$\limsup_{n\to\infty}n\lambda_{\text{max}}(X^\top X)^{-1}<\infty$ where $\lambda_{\text{max}}(A)$ is the largest eigenvalue of $A$. 
\end{assumption}
Assumptions \ref{A1},\ref{A2} and \ref{A3} imply that for each $n$, the matrix $I_{n}-\rho W$ is non-singular and $|I_{n}-\rho W|$ is bounded uniformly in $\rho$ in a closed interval contained in $(-1/\omega_{min},1/\omega_{max})$. This ensures that there exists a consistent root for the system of  likelihood equations in the interior of the parameter space. The full rankness in Assumption \ref{A4} ensures that the term $\lim_{n\to\infty}\frac{1}{n}(X^{\top}X)^{-1}$ that appears in the limiting covariance matrix of $\hat{\beta}$ is well defined. The uniform boundedness of the maximum eigenvalue of $(X^{\top}X)^{-1}$ implies that $\lim_{n\to\infty}\frac{1}{n}\text{Tr}(X^{\top}X)^{-1}=0$ ensuring the consistency of the MLE of $\beta$ and $\sigma^2$. Moreover, since the underlying likelihood function is based on the assumption of normality of the errors, it satisfies the usual Cramer-regularity conditions ensuring the asymptotic efficiency of the MLE. The detailed discussions on various regularity conditions for asymptotic efficiency can be found in \citep{lehmann2006theory,van2000asymptotic}. 
\par Note that since the determinant of a matrix is the product of its eigenvalues, it is straightforward to observe that
\begin{equation*}
    |I_{n}-\rho W|=\prod_{i=1}^{n}(1-\rho \omega_{i}). 
\end{equation*}
Hence $I_{n}-\rho W$ is non-singular if and only if $(1-\rho \omega_{i})\neq 0$ for $i=1,2,...,n$. Moreover, if $W$ is row-normalized, it holds that $|\omega_{i}|\leq 1$. Hence, in such a case, Assumption \ref{A3} implies the existence of a consistent MLE for $\rho$ in a closed set containing $(-1,1)$. Thus, Assumptions \ref{A2} and \ref{A3} can be replaced by the following stronger and easily verifiable assumptions that are found more commonly in the literature. See for instance, \cite{anselin1988spatial}(Chapter 3).  
\begin{assumption}\label{A5}
    The matrix $W$ is row normalized. 
\end{assumption}
\begin{assumption}\label{A6}
    The spatial autocorrelation parameter satisfies $|\rho|<1$. 
\end{assumption}
Assumptions \ref{A1}-\ref{A4} or equivalently Assumptions \ref{A1}, \ref{A4}, \ref{A5} and \ref{A6} imply that the process $\{Y_{n},n\geq 1\}$ (conditional on $X$) is a locally covariant random field in the sense of \cite{smith1980central}. The central limit theorem for such random fields can then be employed to establish that \begin{equation*}
  n^{1/2}(\hat{\theta}-\theta_{0})\xrightarrow{d} N_{p+2}\left(0,\mathcal{I}(\theta_{0})^{-1}\right)  
\end{equation*}
as $n\to\infty$, where $\mathcal{I}(\theta_{0})$ denotes the Fisher information matrix evaluated at $\theta_{0}$. 
\section{Appendix: Proofs} \label{App:B}
\subsection{Proof of Theorem \ref{Th:1}}
The central limit theorem for locally covariant random fields along with the Cramer-Wald device implies that
\begin{equation}\label{eq:A1}
n^{-1/2}
\begin{pmatrix}
U_{n,\rho}\\
U_{n,\sigma^2}\\ U_{n,\mathcal{S},\beta}
\end{pmatrix}\xrightarrow{d}N_{|\mathcal{S}|+2}
\left[\begin{pmatrix}0\\
0 \\
O_{|\mathcal{S}|\times 1}
\end{pmatrix},\mathcal{I}_{\mathcal{S}}\right]
\end{equation}
under $P_{\theta_{0}}$ as $n\to\infty$. Let $\theta_{n}=(\rho,\sigma^2,\beta_{n})$ denote the parameter vector under local misspecification (\ref{2.6}). Let $\theta_{0,S}=(\rho_{0},\sigma_{0}^2,\Pi_{\mathcal{S}}\beta_{0})$ and $\theta_{n,S}=(\rho_{0},\sigma_{0}^2,\Pi_{\mathcal{S}}\beta_{n})$ denote the subvectors of $\theta_{0}$ and $\theta_{n}$ respectively under a subset model $\mathcal{S}$. Let $l(\theta|Y,X,W)=\exp(L(\theta|Y,X,W))$ denote the likelihood where $L(\theta|Y,X,W)$ denotes the (conditional) log-likelihood function. Since the likelihood is twice differentiable with bounded third derivatives on $\Theta$, the Taylor's expansion of second order in a neighborhood of $\theta_{0,\mathcal{S}}$ implies that 
\begin{equation}\label{eq:A2}
\log \frac{l(\theta_{n,\mathcal{S}}|Y,X,W)}{l(\theta_{0,\mathcal{S}}|Y,X,W)}=n^{-1/2}\left(0,0,\Pi_{\mathcal{S}}\delta\right)^{\top}\begin{pmatrix}
U_{n,\rho}\\
U_{n,\sigma^2}\\ U_{n,\mathcal{S},\beta}
\end{pmatrix}-\frac{1}{2}\left(0,0,\Pi_{\mathcal{S}}\delta\right)^{\top}\mathcal{I}_{n}(\theta_{0,\mathcal{S}})\left(0,0,\Pi_{\mathcal{S}}\delta\right)+o_{P_{\theta_{0,\mathcal{S}}}}(1),
\end{equation}
where $\mathcal{I}_{n}(\theta_{0,\mathcal{S}})=\Pi_{\mathcal{S}}I_{n}(\theta_{0})\Pi_{\mathcal{S}}^{\top}$ denotes the submatrix of the empirical Fisher information matrix with respect to a subset model $\mathcal{S}$. Equation (\eqref{eq:A2}) implies that the log-likelihood ratio is locally asymptotically normal (LAN) in the sense of \cite{lecam1970assumptions}. 
Moreover, probability measures $P_{\theta_{0,\mathcal{S}}}$ and $P_{\theta_{n,\mathcal{S}}}$ are mutually absolutely continuous (contiguous). Hence, the expansion of the log-likelihood ratio under the fixed probability model $P_{\theta_{0,\mathcal{S}}}$ as claimed in Equation (\ref{eq:A2}) also holds true under locally misspecified probability model $P_{\theta_{n,\mathcal{S}}}$. LeCam's LAN lemma (\cite{le2000asymptotics}, Chapter 5), (\cite{van2000asymptotic}, Chapter 7) along with Equations (\ref{eq:A1}) and (\ref{eq:A2}) lead to the first statement of the theorem. 
\par Towards proving the second statement, owing to the Cramer-regularity of the likelihood, it can be shown that the sequence $A_{n,\mathcal{S}}=n^{1/2}(\hat{\theta}_{\mathcal{S}}-\theta_{0,\mathcal{S}})$ is asymptotically equivalent to the sequence $B_{n,\mathcal{S}}=\mathcal{I}_{n}(\theta_{0,\mathcal{S}})^{-1}n^{-1/2}\left(U_{n,\rho}, U_{n,\sigma^2}, U_{n,\mathcal{S},\beta}\right)^{\top}
$ 
under $P_{\theta_{0,\mathcal{S}}}$. Since the locally misspecified sequence $P_{\theta_{n,\mathcal{S}}}$ is contiguous to $P_{\theta_{0,\mathcal{S}}}$, the sequences $A_{n,\mathcal{S}}$ and $B_{n,\mathcal{S}}$ are asymptotically equivalent even under $P_{\theta_{n,\mathcal{S}}}$. The asymptotic normality of $B_{n,\mathcal{S}}$ under $P_{\theta_{n,\mathcal{S}}}$ is established by the first statement of the theorem. Invoke Slutsky's lemma along with the Cramer-Wald device to arrive at the desired second statement of the theorem. 
\subsection{Proof of Theorem \ref{Th:2}}
Since $\mu_{W}$ is a differentiable function on $\Theta$, the desired theorem is an immediate consequence of the delta method (\cite{van2000asymptotic}, Theorem 3.1) applied to Theorem \ref{Th:1}. 
\subsection{Proof of Theorem \ref{Th:3}}
Setting $\Pi_{S}=I_{p}$, the subset model $\mathcal{S}$ corresponds to the wide model with the parameter vector $\theta=(\rho,\sigma^2,\beta)$ for which case, the first statement of Theorem \ref{Th:1} implies that 
\begin{equation}\label{eq:A3}
n^{-1/2}
\begin{pmatrix}
U_{n,\rho}\\
U_{n,\sigma^2}\\ U_{n,\beta}
\end{pmatrix}\xrightarrow{d}N_{p+2}
\left[\begin{pmatrix}\mathcal{I}_{\rho,\beta}\delta\\
0 \\
\mathcal{I}_{\beta,\beta}\delta
\end{pmatrix},\mathcal{I}(\theta_{0})\right]
\end{equation}
under $P_{\theta_{n}}$. The contiguity of $P_{\theta_{n}}$ with $P_{\theta_{0}}$ along with Cramer-regularity of the likelihood implies that the sequences $A_{n}=n^{1/2}(\hat{\theta}-\theta_{0})$ and $B_{n}=\mathcal{I}_{n}(\theta_{0})^{-1}n^{-1/2}\left(U_{n,\rho},U_{n,\sigma^2},U_{n,\beta}\right)^{\top}$ are asymptotically equivalent under $P_{\theta_{n}}$. Hence Equation (\ref{eq:A3}) along with Slutsky's lemma shows that the sequence $A_{n}$ is asymptotically $(p+2)$-variate normal with the mean vector $M_{\delta}=\mathcal{I}(\theta_{0})^{-1}\left[\mathcal{I}_{\rho,\beta}\delta~~0~~\mathcal{I}_{\beta,\beta}\delta\right]^{\top}$ and the covariance matrix $\mathcal{I}(\theta_{0})^{-1}$. It is straightforward to observe that \begin{equation*}\begin{split}M_{\delta}&=\mathcal{I}(\theta_{0})^{-1}\begin{pmatrix}\mathcal{I}_{\rho,\beta}\delta\\
0 \\
\mathcal{I}_{\beta,\beta}\delta\end{pmatrix}\\
&=\begin{pmatrix}
\mathcal{I}^{\rho,\rho} & \mathcal{I}^{\rho,\sigma^{2}} & \mathcal{I}^{\rho,\beta}\\
\mathcal{I}^{\sigma^{2},\rho} & \mathcal{I}^{\sigma^{2},\sigma^{2}} & O_{1\times p} \\
\mathcal{I}^{\beta,\rho} & O_{p\times 1}  & \mathcal{I}^{\beta,\beta}
\end{pmatrix}\begin{pmatrix}\mathcal{I}_{\rho,\beta}\delta\\
0 \\
\mathcal{I}_{\beta,\beta}\delta\end{pmatrix}\\
&=\begin{pmatrix}0\\
0 \\
\delta\end{pmatrix},\end{split}\end{equation*} where in the last step, we use the expressions for the blocks of inverse of a partitioned matrix coupled with routine matrix manipulations. See, for instance, \cite{horn2012matrix} (Chapter 3). Now apply Cramer-Wald device to arrive the desired statement of the theorem. 
\subsection{Proof of Theorem \ref{Th:4}}
Let $A_{n,\mathcal{S}}=\left(A_{n,\mathcal{S},\rho},A_{n,\mathcal{S},\sigma^{2}},A_{n,\mathcal{S},\beta}\right)^{\top}$ denote the sequence $A_{n,\mathcal{S}}$ partitioned according to $(\rho,\sigma^2,\beta)$ and in that order. Taylor's expansion of first order implies
\begin{equation}\label{eq:A4}
\begin{split}
n^{1/2}\left(\hat{\mu}_{W,i,\mathcal{S}}-\mu_{W,i}^{*}\right)&=\frac{\partial\mu_{W,i,\mathcal{S}}}{\partial\rho}A_{n,\mathcal{S},\rho}+\frac{\partial\mu_{W,i,\mathcal{S}}}{\partial\sigma^{2}}A_{n,\mathcal{S},\sigma^{2}}+\left(\frac{\partial\mu_{W,i,\mathcal{S}}}{\partial\beta_{\mathcal{S}}}\right)^{\top}A_{n,\mathcal{S},\beta}-\left(\frac{\partial\mu_{W,i,\mathcal{S}}}{\partial\beta_{0}}\right)^{\top}\delta+o_{P_{\theta_{n,\mathcal{S}}}}(1)\\
&=A_{n,\mathcal{S},\rho}\sum_{j=1}^{n}w_{ij}y_{j}+x_{i}^{\top}A_{n,\mathcal{S},\beta}-x_{i}^{\top}\delta+o_{P_{\theta_{n,\mathcal{S}}}}(1).
\end{split}
\end{equation}
Note that $A_{n,\mathcal{S}}=n^{1/2}(\hat{\theta}_{\mathcal{S}}-\theta_{0,\mathcal{S}})$ is asymptotically equivalent to the sequence ~\\$B_{n,\mathcal{S}}=\mathcal{I}_{n}(\theta_{0,\mathcal{S}})^{-1}n^{-1/2}\left(U_{n,\rho}, U_{n,\sigma^2}, U_{n,\mathcal{S},\beta}\right)^{\top}
$ 
under $P_{\theta_{n,\mathcal{S}}}$. Hence, the first statement of Theorem \ref{Th:2} implies that there exist random variables $V_{\rho}\sim N_{1}\left(0,\mathcal{I}_{\rho,\rho}\right)$ and $V_{\beta}\sim N_{p}\left(O_{p\times 1},\mathcal{I}_{\beta,\beta}\right)$ such that
\begin{equation*}
\begin{pmatrix}
A_{n,S,\rho}\\
A_{n,S,\beta}
\end{pmatrix}
=\begin{pmatrix}
\mathcal{I}_{\mathcal{S}}^{\rho,\rho} & \mathcal{I}_{\mathcal{S}}^{\rho,\beta} \\
\mathcal{I}_{\mathcal{S}}^{\beta,\rho}  & \mathcal{I}_{\mathcal{S}}^{\beta,\beta}
\end{pmatrix}
\begin{pmatrix}
\mathcal{I}_{\rho,\beta}\delta+V_{\rho}\\
\Pi_{\mathcal{S}}\mathcal{I}_{\beta,\beta}\delta+\Pi_{\mathcal{S}}V_{\beta}
\end{pmatrix}.
\end{equation*} 
Hence 
\begin{equation*}
A_{n,S,\rho}=\mathcal{I}_{\mathcal{S}}^{\rho,\rho}\mathcal{I}_{\rho,\beta}\delta+\mathcal{I}_{\mathcal{S}}^{\rho,\beta}\Pi_{\mathcal{S}}\mathcal{I}_{\beta,\beta}\delta+\mathcal{I}_{\mathcal{S}}^{\rho,\rho}V_{\rho}+\mathcal{I}_{\mathcal{S}}^{\rho,\beta}\Pi_{\mathcal{S}}V_{\beta},
\end{equation*}
and 
\begin{equation*}
A_{n,S,\beta}=\mathcal{I}_{\mathcal{S}}^{\rho,\beta}\left(\mathcal{I}_{\rho,\beta}\delta+V_{\rho}\right)+\mathcal{I}_{\mathcal{S}}^{\beta,\beta}\left(\Pi_{\mathcal{S}}\mathcal{I}_{\beta,\beta}\delta+\Pi_{\mathcal{S}}V_{\beta}\right).
\end{equation*}
But
\begin{align*}
\mathcal{I}_{\mathcal{S}}^{\beta,\beta}&=Q_{\mathcal{S}}, \notag \\
\mathcal{I}_{\mathcal{S}}^{\rho,\beta}&=-\mathcal{I}_{\rho,\rho}^{-1}\mathcal{I}_{\rho,\beta}\Pi_{\mathcal{S}}^{\top}Q_{\mathcal{S}},\notag \\
\mathcal{I}_{\mathcal{S}}^{\rho,\rho}&=\mathcal{I}_{\rho,\rho}^{-1}+\mathcal{I}_{\rho,\rho}^{-1}\mathcal{I}_{\rho,\beta}\Pi_{\mathcal{S}}^{\top}Q_{\mathcal{S}}\Pi_{\mathcal{S}}\mathcal{I}_{\beta,\rho}\mathcal{I}_{\rho,\rho}^{-1} \notag \\
G_{\mathcal{S}}&=\Pi_{\mathcal{S}}^{\top}Q_{\mathcal{S}}\Pi_{\mathcal{S}}Q^{-1}.
\end{align*}
Hence
\begin{equation}\label{eq:A5}
A_{n,S,\rho}=\mathcal{I}_{\rho,\rho}^{-1}\mathcal{I}_{\rho,\beta}\left(I_{p}-G_{\mathcal{S}}\right)\delta+\mathcal{I}_{\mathcal{S}}^{\rho,\rho}V_{\rho}-\mathcal{I}_{\rho,\rho}^{-1}\mathcal{I}_{\rho,\beta}G_{\mathcal{S}}(C-\delta),
\end{equation}
and 
\begin{equation}\label{eq:A6}
\begin{split}
A_{n,S,\beta}&=Q_{\mathcal{S}}\Pi_{\mathcal{S}}Q^{-1}\delta+Q_{\mathcal{S}}\Pi_{\mathcal{S}}Q^{-1}(C-\delta)
\\&=Q_{\mathcal{S}}\Pi_{\mathcal{S}}Q^{-1}C.
\end{split}
\end{equation}
where $C$ is the matrix as given in the statement of the theorem. Use Equations (\ref{eq:A5}) and (\ref{eq:A6}) in (\ref{eq:A4}) to arrive at the desired expression. 
\par We now establish the independence of the two random variables appearing in $\Lambda_{W,i,\mathcal{S}}$. Observe that
\begin{equation*}
\begin{pmatrix}
V_{\rho}\\
V_{\beta}-\mathcal{I}_{\beta,\rho}\mathcal{I}_{\rho,\rho}^{-1}V_{\rho}
\end{pmatrix}=
\begin{pmatrix}
 1   & O_{1\times p}\\
 -\mathcal{I}_{\beta,\rho}\mathcal{I}_{\rho,\rho}^{-1} & I_{p}
\end{pmatrix}\begin{pmatrix}
V_{\rho}\\V_{\beta}
\end{pmatrix}.
\end{equation*}
Hence, the random vector $(V_{\rho},C-\delta)^{\top}$ is $(p+1)$ variate normal with the block diagonal covariance matrix 
\begin{equation*}
\begin{pmatrix}
\mathcal{I}_{\rho,\rho}& O_{1\times p}\\
O_{p\times 1} & Q
\end{pmatrix}
\end{equation*}
implying the independence of $V_{\rho}$ and $C-\delta$. Note that the random variable $\Lambda_{W,i}$ is a measurable transformation of $V_{\rho}$ while $\omega_{W,i}^{\top}(\delta-G_{\mathcal{S}}C)$ is measurable transformation of a linear combination of $V_{\rho}$ and $C-\delta$. Hence the independence of these terms is immediate due to the independence of $V_{\rho}$ and $C-\delta$. 
\subsection{Proof of Theorem \ref{Th:5}}
Using the expression for the risk $R_{W,i,\mathcal{S}}$ in Equation (\ref{AMSE1}), we get
\begin{equation*}
\begin{split}
\eta_{W,\psi,\mathcal{S}}&=\int_{\mathbb{R}^{p}} \omega_{W}(z)^{\top}\left[\left(I_{p}-G_{\mathcal{S}}\right)\delta\delta^{\top}\left(I_{p}-G_{\mathcal{S}}\right)^{\top}+G_{\mathcal{S}}QG_{\mathcal{S}}^{\top}\right]\omega_{W}(z)dF_{\psi}(z)+\int_{\mathbb{R}^{p}} \left(\frac{\partial\mu_{W(z)}}{\partial\rho}\right)^{2}I_{\rho,\rho}^{-1}dF_{\psi}(z)\\
&=\text{Tr}\left(\left(I_{p}-G_{\mathcal{S}}\right)\delta\delta^{\top}\left(I_{p}-G_{\mathcal{S}}\right)^{\top}K_{W,\psi}\right)+\text{Tr}\left(G_{\mathcal{S}}QG_{\mathcal{S}}^{\top}K_{W,\psi}\right)+\int_{\mathbb{R}^{p}} \left(\frac{\partial\mu_{W(z)}}{\partial\rho}\right)^{2}I_{\rho,\rho}^{-1}dF_{\psi}(z),
\end{split}
\end{equation*}
where
\begin{equation*}
\begin{split}
K_{W,\psi}&=\int_{\mathbb{R}^{p}} \omega_{W}(z)\omega_{W}(z)^{\top}dF_{\psi}(z)\\
&=\int_{\mathbb{R}^{p}}\left(\mathcal{I}_{\beta,\rho}\mathcal{I}_{\rho,\rho}^{-1}\frac{\partial\mu_{W}(z)}{\partial\rho}-\frac{\partial\mu_{W}(z)}{\partial\beta}\right)\left(\mathcal{I}_{\beta,\rho}\mathcal{I}_{\rho,\rho}^{-1}\frac{\partial\mu_{W}(z)}{\partial\rho}-\frac{\partial\mu_{W}(z)}{\partial\beta}\right)^{\top}dF_{\psi}(z)\\
&=\mathcal{I}_{\beta,\rho}\mathcal{I}_{\rho,\rho}^{-1}H_{W,\psi,\rho,\rho}\mathcal{I}_{\rho,\rho}^{-1}\mathcal{I}_{\rho,\beta}-2\mathcal{I}_{\beta,\rho}\mathcal{I}_{\rho,\rho}^{-1}H_{W,\psi,\rho,\beta}+H_{W,\psi,\beta,\beta}.
\end{split}
\end{equation*}
Hence the proof.
\end{document}